\documentclass[iop]{emulateapj}

\usepackage{graphicx}
\usepackage{amsmath}
\usepackage{natbib}
\usepackage{mathptmx}
\usepackage{enumerate}
\usepackage{apjfonts}
\usepackage{times}
\usepackage[colorlinks=true,urlcolor=blue,citecolor=blue,linkcolor=blue]{hyperref}
\usepackage{aas_macros}
\usepackage[position=t,singlelinecheck=on,font={rm,bf,up},font=large]{subfig}
\usepackage{capt-of}

\makeatletter
\long\def\frontmatter@title@above{
  \vspace*{-17mm}\vspace*{\headheight}
   \hspace{-3mm}{\sc The Astrophysical Journal Supplement Series}, 229:11 (11pp), 2017\\
   \vspace*{4mm}{\footnotesize {\sc Preprint typeset using \LaTeX\ style emulateapj}}
  \par\vspace*{-\baselineskip}\vspace{6mm}
  }

\makeatother

\RequirePackage{color}

\bibpunct{(}{)}{;}{a}{}{,}

\usepackage{pdfcomment,acronym}

\shorttitle{Low-lying fields in the solar chromosphere}
\shortauthors{Jafarzadeh et al.}

\begin{document}

\title{Slender Ca\,{\sc ii}\,H Fibrils Mapping Magnetic Fields in the Low Solar Chromosphere}

\author{S.~Jafarzadeh\hyperlink{}{\altaffilmark{1}}}
\author{R.~J.~Rutten\hyperlink{}{\altaffilmark{1}}}
\author{S.~K.~Solanki\hyperlink{}{\altaffilmark{2,3}}}
\author{T.~Wiegelmann\hyperlink{}{\altaffilmark{2}}}
\author{T.~L.~Riethm\"{u}ller\hyperlink{}{\altaffilmark{2}}}
\author{M.~van~Noort\hyperlink{}{\altaffilmark{2}}}
\author{M.~Szydlarski\hyperlink{}{\altaffilmark{1}}}
\author{J.~Blanco~Rodr\'{i}guez\hyperlink{}{\altaffilmark{4}}}
\author{P.~Barthol\hyperlink{}{\altaffilmark{2}}}
\author{J.~C.~del~Toro~Iniesta\hyperlink{}{\altaffilmark{5}}}
\author{A.~Gandorfer\hyperlink{}{\altaffilmark{2}}}
\author{L.~Gizon\hyperlink{}{\altaffilmark{2,6}}}
\author{J.~Hirzberger\hyperlink{}{\altaffilmark{2}}}
\author{M.~Kn\"{o}lker\hyperlink{}{\altaffilmark{7,}\altaffilmark{10}}}
\author{V.~Mart\'{i}nez~Pillet\hyperlink{}{\altaffilmark{8}}}
\author{D.~Orozco~Su\'{a}rez\hyperlink{}{\altaffilmark{5}}}
\author{W.~Schmidt\hyperlink{}{\altaffilmark{9}}}

\affil{\altaffilmark{1}\hspace{0.2em}Institute of Theoretical Astrophysics, University of Oslo, P.O. Box 1029 Blindern, N-0315 Oslo, Norway; \href{mailto:shahin.jafarzadeh@astro.uio.no}{shahin.jafarzadeh@astro.uio.no}\\
\altaffilmark{2}\hspace{0.2em}Max Planck Institute for Solar System Research, Justus-von-Liebig-Weg 3, 37077 G\"{o}ttingen, Germany\\
\altaffilmark{3}\hspace{0.2em}School of Space Research, Kyung Hee University, Yongin, Gyeonggi 446-701, Republic of Korea\\
\altaffilmark{4}\hspace{0.2em}Grupo de Astronom\'{\i}a y Ciencias del Espacio, Universidad de Valencia, 46980 Paterna, Valencia, Spain\\
\altaffilmark{5}\hspace{0.2em}Instituto de Astrof\'{i}sica de Andaluc\'{i}a (CSIC), Apartado de Correos 3004, E-18080 Granada, Spain\\
\altaffilmark{6}\hspace{0.2em}Institut f\"ur Astrophysik, Georg-August-Universit\"at G\"ottingen, Friedrich-Hund-Platz 1, 37077 G\"ottingen, Germany\\
\altaffilmark{7}\hspace{0.2em}High Altitude Observatory, National Center for Atmospheric Research, \footnote{The National Center for Atmospheric Research is sponsored by the National Science Foundation.} P.O. Box 3000, Boulder, CO 80307-3000, USA\\
\altaffilmark{8}\hspace{0.2em}National Solar Observatory, 3665 Discovery Drive, Boulder, CO 80303, USA\\
\altaffilmark{9}\hspace{0.2em}Kiepenheuer-Institut f\"{u}r Sonnenphysik, Sch\"{o}neckstr. 6, D-79104 Freiburg, Germany}

\altaffiltext{10}{The National Center for Atmospheric Research is sponsored by the National Science Foundation.}

\begin{abstract} A dense forest of slender bright fibrils near a small solar active region is seen in high-quality narrowband Ca\,{\sc ii}\,H images from the SuFI instrument onboard the {\sc Sunrise} balloon-borne solar observatory.
The orientation of these slender Ca\,{\sc ii}\,H fibrils (SCF) overlaps with the magnetic field configuration in the low solar chromosphere derived by magnetostatic extrapolation of the photospheric field observed with {\sc Sunrise}/IMaX and SDO/HMI. 
In addition, many observed SCFs are qualitatively aligned with small-scale loops computed from a novel inversion approach based on
best-fit numerical MHD simulation.   
Such loops are organized in canopy-like arches over quiet areas that differ in height depending on the field strength near their roots.

\vspace{1mm}
\end{abstract}


\section{Introduction}
\label{sec:intro}

Large-scale solar magnetism is thought to be generated by
dynamo processes in the solar interior 
(reviews by
\citealt{Choudhuri2003,Ossendrijver2003,Charbonneau2005,Charbonneau2010}).
The interior magnetic fields rise through the convection zone driven by their
buoyancy, emerge at the surface, and extend through the entire solar
atmosphere \citep{Moreno-Insertis1986,Schussler1994,Caligari1995}. 
They manifest themselves at the photospheric surface on all spatial
scales, expand with height, and re-orient with a range of inclination
angles depending on their larger-scale configurations and mutual
interactions
\citep[e.g.,][]{Stenflo1989,Solanki1993,Solanki2001,Solanki2006,deWijn2009}.

In addition, small-scale magnetic fields are thought to be produced by
small-scale turbulent dynamo action close to the surface
(\citealt{Voegler2007,Danilovic2010a,Stenflo2012,Rempel2014}; see also
\citealt{Borrero2015} for a review.)

Magnetic canopies are particular field configurations in which nearly horizontal field lines lie over a region with weaker (or no) field. They are produced when the field in magnetic elements, pores, or sunspots bends over as these features expand. These horizontal field lines can either  return to the surface, connecting magnetic
surface features of opposite polarities, or expand laterally until
they meet same-polarity fields from other surface features (e.g.,
\citealt{Gabriel1976,Jones1985,Solanki1990,Bray1991}; review by
\citealt{Wedemeyer-Bohm2009}). 
Canopies may occur at different atmospheric heights.
Estimations range between the mid photosphere and high chromosphere
depending on the size and field strength of the concentrations at the
surface and/or the distances between
them~\citep[e.g.,][]{Spruit1981b,Giovanelli1982,Jones1982,Roberts1990,Buente1993,Bruls1995,Solanki1991,Zhang2000}.  
A numerical simulation by \citet{Rosenthal2002} suggested canopy
heights ranging over 800-1600~km for network and internetwork areas, with the value depending strongly on the thermodynamic properties of the atmosphere inside and outside the magnetic feature~\citep{Solanki1990}.

Direct observation of magnetic fields at chromospheric heights are
rather challenging and consequently rare (see \citealt{Lagg2015}).
It is generally believed that the long fibrils visible in the core of
H$\alpha$ wherever there is some magnetic activity are mapping fields
overlying quieter internetwork regions (e.g.,
\citealt{Wiegelmann2008}) 
in the form of supergranulation cell-covering canopies.
Comparisons between elongated structuring and field topography
in the upper chromosphere have suggested that fibrils outline magnetic
fields at these heights, from Stokes inversions of the Ca~{\sc
ii}~854.2~nm line by \citet{de-laCruz2011} and \citet{Rouppe2013}, of
the He~{\sc i}~1083~nm line by \citet{Schad2013}. 
The most detailed modeling test of this assumption was done
by \citet{Leenaarts2015} 
using a state-of-the-art MHD simulation, concluding that the
synthetic fibrils map field lines fairly well during their
start but not at subsequent wave modulation.

\citet{Schrijver2003} 
have argued that there should be ubiquitous magnetic loops
returning to the surface at much smaller scales than network or
supergranulation cells. 
Such shorter-extent lower-lying canopies in active regions have not been found so far. Although, in the quiet Sun, \citet{Wiegelmann2010} and \citet{Wiegelmann2013} showed the presence of short, low-lying, and highly dynamic loops from magnetic field extrapolations. 

This study addresses this issue using high-quality image sequences
from the {\sc Sunrise} balloon-borne solar observatory during its
second flight \citep{Solanki2017}.  
The target was a growing active region containing a leading spot (not observed) and a group of pores near disk center.
In a high-resolution image sequence obtained with the narrowband
Ca\,{\sc ii}\,H filter in the {\sc Sunrise}/SuFI imager we noticed
very thin and long bright features, which we call slender
Ca\,{\sc ii}\,H fibrils (SCF) henceforth. 
They appear to emanate from the more active areas in the field of view
that contain pores and dense small strong-field magnetic concentrations.
The physical properties of these SCFs have been investigated by \citet{Gafeira2017a}. In addition, \citet{Jafarzadeh2017b} and \citet{Gafeira2017b} studied transverse and sausage-mode oscillations in the {\sc Sunrise}/SuFI SCFs and provided propagation speeds of both types of waves.

The on-disk SCFs represent a rather new phenomenon.
Similar fibrils were seen earlier in high-resolution Ca\,{\sc
ii}\,K and Ca\,{\sc ii}\,H images from the Swedish 1-m Solar Telescope
\citep{Pietarila2009,Henriques2013} and in {\sc Sunrise}-II Mg\,{\sc
ii}\,k and Ca\,{\sc ii}\,H images
\citep{Riethmuller2013c,Danilovic2014}. We note that, although
from the same instrument and the same Ca\,{\sc ii}\,H passband, the
data set under study samples a different solar region and is of higher
quality than those shown by \citet{Riethmuller2013c} and
\citet{Danilovic2014} (i.e., they were treated by an improved
image-reconstruction technique).

Comparable long slender fibrils were also observed near the solar limb
in Ca\,{\sc ii}\,H images from the Dutch Open Telescope and called ``straws''
by \citet{Rutten2006}.
These were then equated to off-limb spicules-II discovered in Ca\,{\sc ii}\,H
image sequences with \textit{Hinode} and identified as Alv\'enic-wave phenomena by
\citet{DePontieu2007c}
with torsion added by
\citet{DePontieu2012}.
Their on-disk counterpart in H$\alpha$ takes the form of rapid
blue-wing excursions (RBE,
\citealt{Langangen2008,Rouppe2009,Sekse2012})
and similar rapid red-wing excursions (RRE,
\citealt{Sekse2013b}).
Their drivers remain unidentified \citep{Pereira2012}.

The {\sc Sunrise} SCFs differ from such spicule-II straws because they are
found near disk center and are extremely thin.
It would be of much interest to compare them to RBEs and RREs in
H$\alpha$ and also to compare them with the onsets of the longer and
wider fibrils observed in the core of H$\alpha$.
Unfortunately, there were no
simultaneous high-resolution observations in these lines of the
{\sc Sunrise} target, so that we cannot study the question whether, and if
so how, the {\sc Sunrise} SCFs relate to spicules-II and/or long H$\alpha$
fibrils.

On the other hand, the {\sc Sunrise} observations do provide high-quality
solar-surface magnetograms from the IMaX instrument
\citep{Martinez-Pillet2011}, while modern field extrapolation
techniques \citep[e.g.,][]{Wiegelmann2015} and yet newer
data-constrained MHD simulation techniques \citep{Riethmuller2017}
permit trustworthy estimation of the actual field configuration above
the surface from these magnetograms.
We employ these techniques here to test the premise that the observed
SCFs are field-aligned and may therefore serve to establish the
magnetic configuration above the surface. 
We concentrate in particular on the question what height these
fibril-marked fields reach.

The organization of this publication
is as follows. 
In the next section, we use a well-documented state-of-the-art
numerical MHD simulation to study magnetic configurations at multiple
heights.  
In Section~\ref{sec:observations} we detail the {\sc Sunrise} observations.
In Section~\ref{sec:modeling} we compare these to magnetic field
extrapolations and MHD simulations defined by the IMaX magnetograms.
We end the study with a brief conclusion.

\section{Field Configuration in a Bifrost Simulation}
\label{sec:bifrost}

In order to illustrate magnetic configurations located at multiple
heights from the photosphere to the corona we visualize the geometry
of magnetic field lines in a 3D snapshot from the well-documented
state-of-the-art simulation with the 3D radiative-MHD code Bifrost \citep{Gudiksen2011}
that has been described and made public by
\citet{Carlsson2016},
to which we refer for more detail.
The same simulation was used by
\citet{Leenaarts2015}.
In brief, the Bifrost code solves the MHD equations on a staggered
grid and includes radiation by solving the radiative transfer
equations along many rays using a short-characteristic method and
multi-group opacities \citep{Nordlund1982,Hayek2010}. 
The chromospheric radiative losses are computed in non-LTE and thermal
conditions at high temperatures are taken into account.  
In this simulation, the hydrogen ionization balance was computed
including non-equilibrium retardation. 

The simulation snapshot used here is dominated by two
opposite-polarity enhanced network patches and samples the entire
atmosphere in a volume of $24\times24$~Mm$^2$ horizontally (with
$504\times504$ grid points giving 48 km resolution), extending from
2.4~Mm below the visible surface (defined as optical depth unity at
500~nm) to 14.4~Mm above it. The vertical dimension is spanned by 496 grid
points with variable sizes changing from 19~km up to a height of 5~Mm
from the bottom of the computational box and then increasing to 100~km
close to the top boundary.
We limit the vertical dimension to 14~Mm in our displays here (see
below). The maximum field strength has an absolute value of 1.9~kG at
the visible solar surface.

To visualize the magnetic configuration we display field lines
through the entire simulation cube that pass through specific
$(x,y,z)$ point locations, which are randomly positioned through a layer at a selected height.
The number of these starting points for field-line tracing is about 2000 and is kept small for
better visibility.
A tracing algorithm then produces field lines following the vector
field from these initial starting points in both directions until they meet a boundary
surface of the simulation volume.
This algorithm so fills the simulation cube with field lines that
chart fields over a wide range of strengths and configurations, with the
common property that they all pass through the selected layer and do so
at random locations.

\begin{figure*}[!thp] \centering
  \includegraphics[width=.98\textwidth, trim = 0 0 0 0, clip]{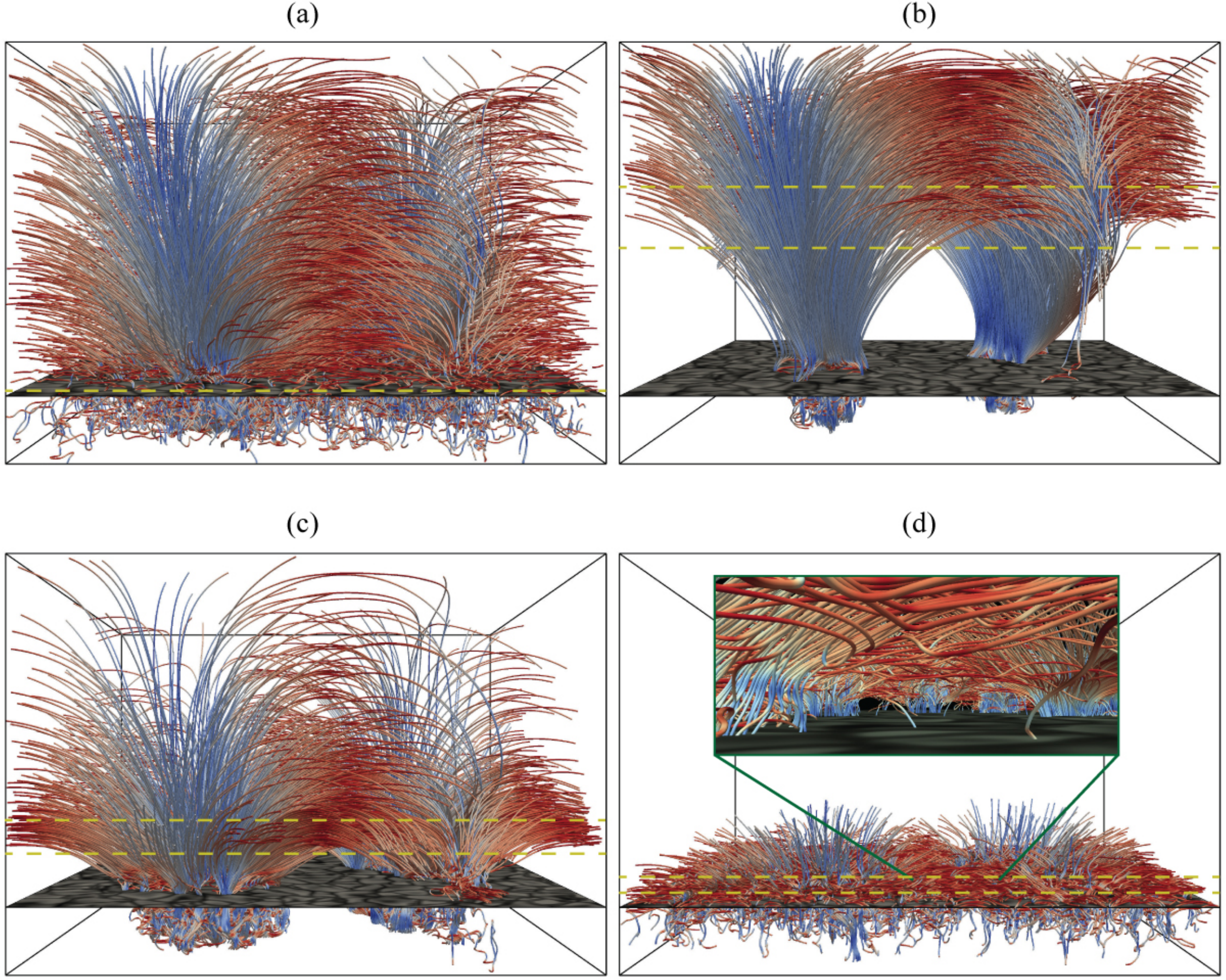}
  \caption[]{Magnetic configuration in the Bifrost simulation in
  perspective view. 
  The volume measures $24\times24$~Mm$^2$ horizontally, 14~Mm
  vertically. 
  The visible surface (optical depth unity at 500~nm) is specified
  with a synthetic continuum image.
  The plotted field lines start at locations that are randomly
  distributed through a thin layer at a selected height and then traced
  upward and downward to chart fields elsewhere (a) throughout the entire simulation volume
  (starting points of the tracing located in a 0--100\,km layer), (b) only those that reach the
  lower corona (starting points between 5000 and 7000\,km), (c) adding
  those that reach the upper chromosphere and transition region
  (starting points 1800--2800\,km), (d) adding those that reach the lower
  chromosphere traced from starting points at 500--1000\,km and with a cut at
  1400~km height for better visibility. 
  The colors represent inclination, from vertical (blue) to horizontal
  (red). 
  A small 700-km high cutout of the region between the two main footpoints is magnified in the inset in (d).}
  \label{fig:bifrost}
\end{figure*}

Selected results are shown in Figure~\ref{fig:bifrost}.
Panel (a) has the starting layer at the surface so that the traced
field lines sample all fields in the simulation volume. We note that all fields in the simulation box originate below the surface, i.e., there are no sagging loops coming into the box from the top or the sides.
By putting the selective starting layer higher, in the low corona,
transition region/upper chromosphere, and low chromosphere
respectively, the partial corresponding field configurations are
mapped in panels (b)--(d).
For better visibility, the field lines in panel (d) are cut at height
1400~km.
The selected starting layers in panels (a)--(d) are marked with the horizontal dashed lines and range between, 0--100\,km, 5000--7000\,km, 1800-2800, and 500--1000\,km, respectively.
We comment that fewer high-lying field lines are seen in panel (c) compared to those in panel (b) because fewer of them are caught when the number of traced field lines is fixed and we start deeper down in the atmosphere.

The height reached by a single field line in Figure~\ref{fig:bifrost}
is usually higher when the field is stronger in the root where it
passes through the surface.
As a result, the figure suggests the presence of a particular type of
magnetic canopy, if we consider fields originating at the surface in a particular type of feature. Depending on the type of footpoint, the height of its lower boundary surface can be quite different. 
For example, panel (b) displays a large volume relatively free of
field extending to large height between and around the two
network patches, with many patch-connecting field lines above it. 
This is because most field lines in panel (b), which selects only
high-reaching ones, are rooted in the two patches of strong-field
opposite-polarity network to gain such height. 
The arch they span above the quiet area between the network patches
represents a high canopy.
This is not a canopy with field-free or high-$\beta$ plasma underneath as
in the more common definitions of magnetic canopy \citep{Solanki1990}, or in the canopy definition as sound-speed and Alfv\'en-speed equality of \citet{Bogdan2003}, but as a dome having unplotted fields inside that originate in less
strong sources and therefore reach less high.

Similarly, the inset in panel (d) shows a small interior part of the
volume which illustrates a low and smaller canopy of this type between
the two network areas that remains within the low chromosphere.
It implies that almost all field lines passing through 500\,km height
connect to stronger surface fields closer to the network patches (blue
stalks to the side) and do not connect the quiet surface containing mainly weak fields separating the strong opposite polarity patches.

The importance  of this display format showing root-strength canopies
with different heights is that they indicate connectivity for actual
observed feature canopies to properties of their source regions. 
For  example, if the  observed canopies outlined by long H$\alpha$
fibrils  reach a  specific  height or  have  a characteristic  minimum
height, then the suggestion from  this simulation rendering is that the
mechanism in the fibril root region that produces the observed
fibrils must operate at a specific  field strength, or only at values
above a particular field strength.

The conclusion from this demonstration is that the existence of many
small low-lying loops in internetwork regions postulated by
\citet{Schrijver2003}
is supported by the Bifrost simulation, but that it also suggests that
there is a hierarchy of canopy-like arched field configurations, which
map field strengths near the field roots at the surface,
and that the height of these arches increases with higher field strengths at the roots, or at least at one of the roots (see \citealt{Wiegelmann2010}).

\vspace{4mm}
\section{Observations from {\sc Sunrise}} \label{sec:observations}
\subsection{Observations and Reduction} 

The primary data set used here was recorded with the narrowband Ca~{\sc
ii}~H filter (with FWHM $\approx0.11$~nm) of the {\sc Sunrise} Filter
Imager (SuFI;~\citealt{Gandorfer2011}) onboard the 1-meter {\sc
Sunrise} balloon-borne solar
observatory~\citep{Solanki2010,Barthol2011,Berkefeld2011} during its
second flight in 2013 ({\sc Sunrise}-II; \citealt{Solanki2017}). 
Our seeing-free image sequence with high spatial and temporal
resolution was collected between 23:39~UT on 2013 June 12 and 00:38~UT
on 2013 June 13. 
The Ca\,{\sc ii}\,H images cover a field of view (FOV) of
($15\times38$)~arcsec$^2$ that covered part of the following polarity magnetic features of NOAA AR 11768 (including a few small pores and plages) near solar disc center
(cosine of the heliocentric angle $\mu \approx 0.93$). 
These high-quality data were corrected for wavefront aberrations by
multi-frame blind deconvolution (MFBD;~\citealt{vanNoort2005}).

We note that several other Ca\,{\sc ii}\,H image sequences of various
solar regions were obtained during the {\sc Sunrise} observations from
2013 June 12 to 2013 June 17 (\citealt{Solanki2017}). 
For comparison with photospheric structures, we also use images
recorded with the {\sc Sunrise}/SuFI 300~nm filter.
In addition, full-Stokes ($I, Q, U$, and $V$) images in the
magnetically sensitive line Fe~{\sc i}~$525.02$~nm were obtained with
the {\sc Sunrise} Imaging Magnetograph eXperiment
(IMaX;~\citealt{Martinez-Pillet2011}).

We use the magnetograms obtained from the {\sc Sunrise}/IMaX images to
extrapolate the magnetic field into the solar chromosphere.
These extrapolations are based on magnetohydrostatic equilibria and have been described by \citet{Wiegelmann2017}. They also employ wider FOV but lower resolution and lower sensitivity 
SDO/HMI data \citep{Pesnell2012,Schou2012}. An earlier application of the magnetohydrostatic equilibrium technique to quiet-Sun observations from the 2009 flight of {\sc Sunrise} ({\sc Sunrise}-I) was made by \citet{Wiegelmann2015}, who also gave a more detailed description of the employed method.

We also compared our Ca\,{\sc ii}\,H images with co-aligned observations
in He~{\sc ii}~30.4~nm from SDO/AIA \citep{Lemen2012}.

\vspace{2mm}
\subsection{{\sc Sunrise} Results}

\begin{figure*}[!thp] \centering
  \includegraphics[width=.98\textwidth, trim = 0 0 0 0, clip]{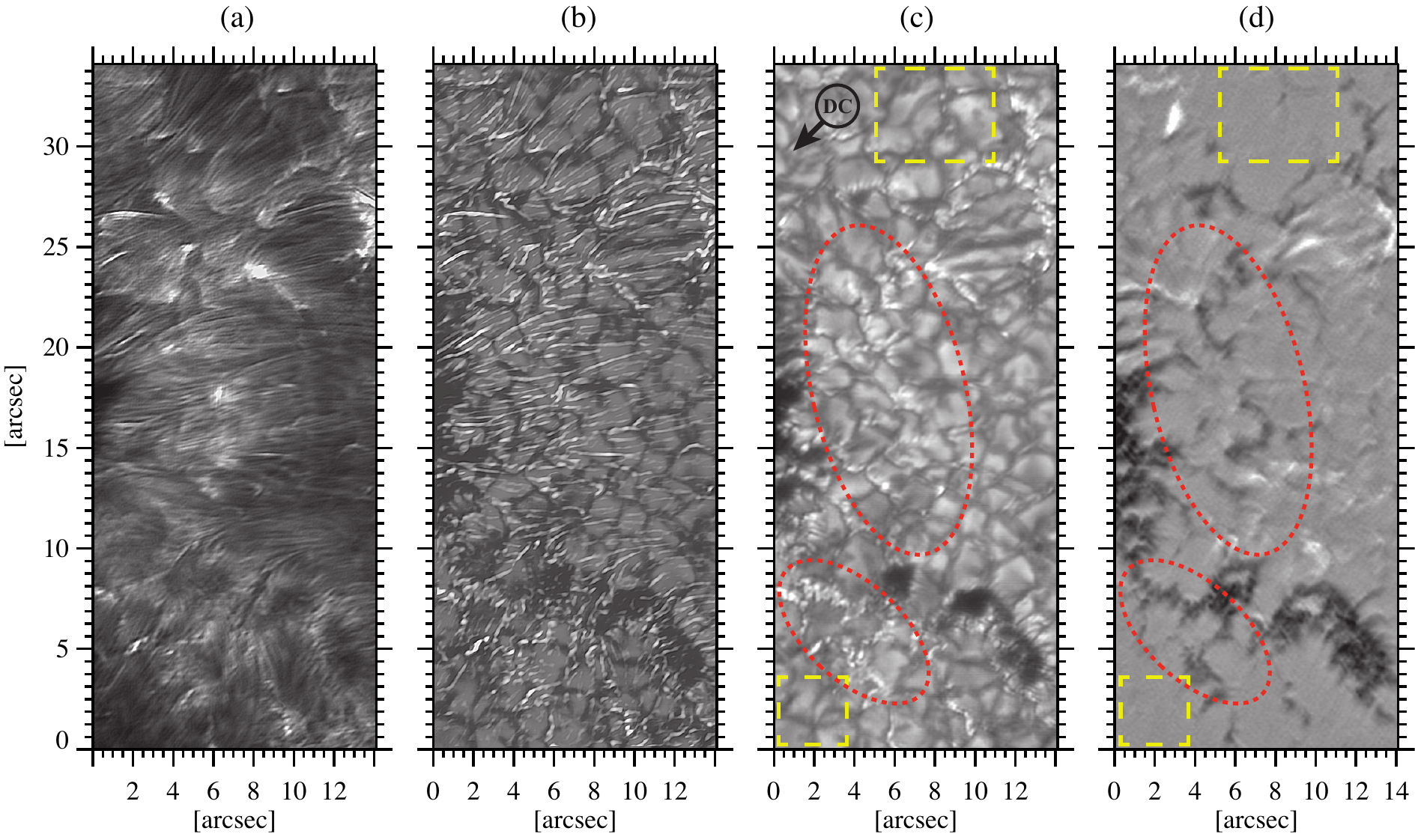}
  \caption[]{Co-aligned simultaneous images from {\sc Sunrise}/SuFI
  (a)--(c) and {\sc Sunrise}/IMax (d).
  Panel (a) is a Ca\,{\sc ii}\,H image showing many SCFs that
  fan out from active areas at left and near the bottom. 
  It has been sharpened for better SCF visibility. 
  Panel (b) shows a superposition of SCF identifications on a darkened
  low-contrast version of the corresponding 300~nm image.
  Panel (c) is the latter at full contrast. 
  Panel (d) is the corresponding Stokes $V$ magnetogram. 
  The red ellipses outline two areas with many magnetic
  concentrations and the yellow boxes two quiet areas. 
  The direction to disk center is specified in panel (c).}
  \label{fig:obs_images}
\end{figure*}

\paragraph{Slender Ca\,{\sc ii}\,H fibrils}
Figure~\ref{fig:obs_images}a shows a Ca\,{\sc ii}\,H image after
sharpening with an unsharp-mask filter to gain better visibility of
the slender fibrils. 
The SCFs in this sequence were identified by restoring the Ca\,{\sc
ii}\,H images in a multi-step procedure described in detail by
\citet{Jafarzadeh2017b}. 
In short, the inhomogeneities in the images due to non-uniform
background solar intensity, noise, and geometric distortions are
minimized using a real-space spatial bandpass filter
\citep{Jafarzadeh2013a} and the SCFs are modeled as bright elongated
structures in the restored images.

Figure~\ref{fig:obs_images}b overplots the detected SCFs corresponding to the
first panel superimposed on a low-contrast version of the co-spatial
and co-temporal 300~nm wider-band image.
Figure~\ref{fig:obs_images}c displays this wide-band image at full
contrast.

The corresponding Stokes $V$ map from IMaX is plotted in Figure~\ref{fig:obs_images}d. 
It is an average over four wavelength positions (at $\pm0.008$~nm and
$\pm0.004$~nm from the Fe~{\sc i}~$525.02$~nm line core) that was
re-scaled and co-aligned with the SuFI images in panels (a)--(c). 
The red ellipses in Figures~\ref{fig:obs_images}c and
\ref{fig:obs_images}d mark two areas with fairly dense network field
concentrations from which many of the fibrils emanate outward with
varying lengths, although many fibrils start already to the left of these features (as seen in the figure), seeming to come from the large pore at the edge of the FOV.
The yellow rectangles outline the quietest areas in the SuFI FOV.
The FOV of the images shown in Figure~\ref{fig:obs_images} is vertically flipped and slightly rotated with respect to the true orientation on the Sun. For the correct orientation, see \citet{Solanki2017}.

We now list the main SCF properties noted from these observations. 
\begin{enumerate} 

\item There are relatively short SCFs radiating out from the large
pore near the mid-left edge of the images. Long SCFs fanning out from this pore are also observed.
The fibrils emanating away from the smaller pores near the
bottom-right corner of Figure~\ref{fig:obs_images}b are shorter than
those from the large pore. 
These are among the shortest in our image sequence.

\item Many long SCFs emanate from the plage near the pores. 
These plage regions
are mostly concentrated in areas to the right side of the large pore
and the lower left side of the small pores that are marked by red
ellipses in Figure~\ref{fig:obs_images}.
While the fibrils starting in the former extend mostly close to the $x$ direction, the latter
are more oriented in the $y$ direction.
A few short SCFs with other angles are also observed in both areas.

\item Inspection of SCF endings on the photospheric 300~nm images
shows that some SCFs seem to have one or both footpoints rooted in
magnetic field concentrations, whereas many others seem to be located
at granules. 
The SCFs emanating from the pores seem to always have one footpoint in
the umbra. 
However, we stress that such footpoint identification is
rather uncertain.

\item The two quiet areas marked by yellow boxes show some 
SCFs but these mostly extend from neighboring magnetic patches.

\end{enumerate}

\paragraph{Height of formation}
We now turn to the question of whether these observations in themselves
permit estimation of SCF formation heights.  
In order to obtain an initial estimate, we have computed
line-depression contribution functions (CF) for the narrowband SuFI
Ca\,{\sc ii}\,H filter using the 1D version of the RH~code of
\cite{Uitenbroek2001}, which solves the radiative transfer and
statistical equilibrium equations in non-LTE and includes partial
redistribution for a given 1D model atmosphere. 
This computation is similar to the one reported in
\citet{Danilovic2014}. 
Our motivation to include and present these rather simplistic
estimates ignoring the actual fine structure of the solar atmosphere
is that they already suggest a bimodal response that likely also affects our
observations.

\begin{figure}[!thp] \centering
  \includegraphics[width=8cm, trim = 0 0 0 0, clip]{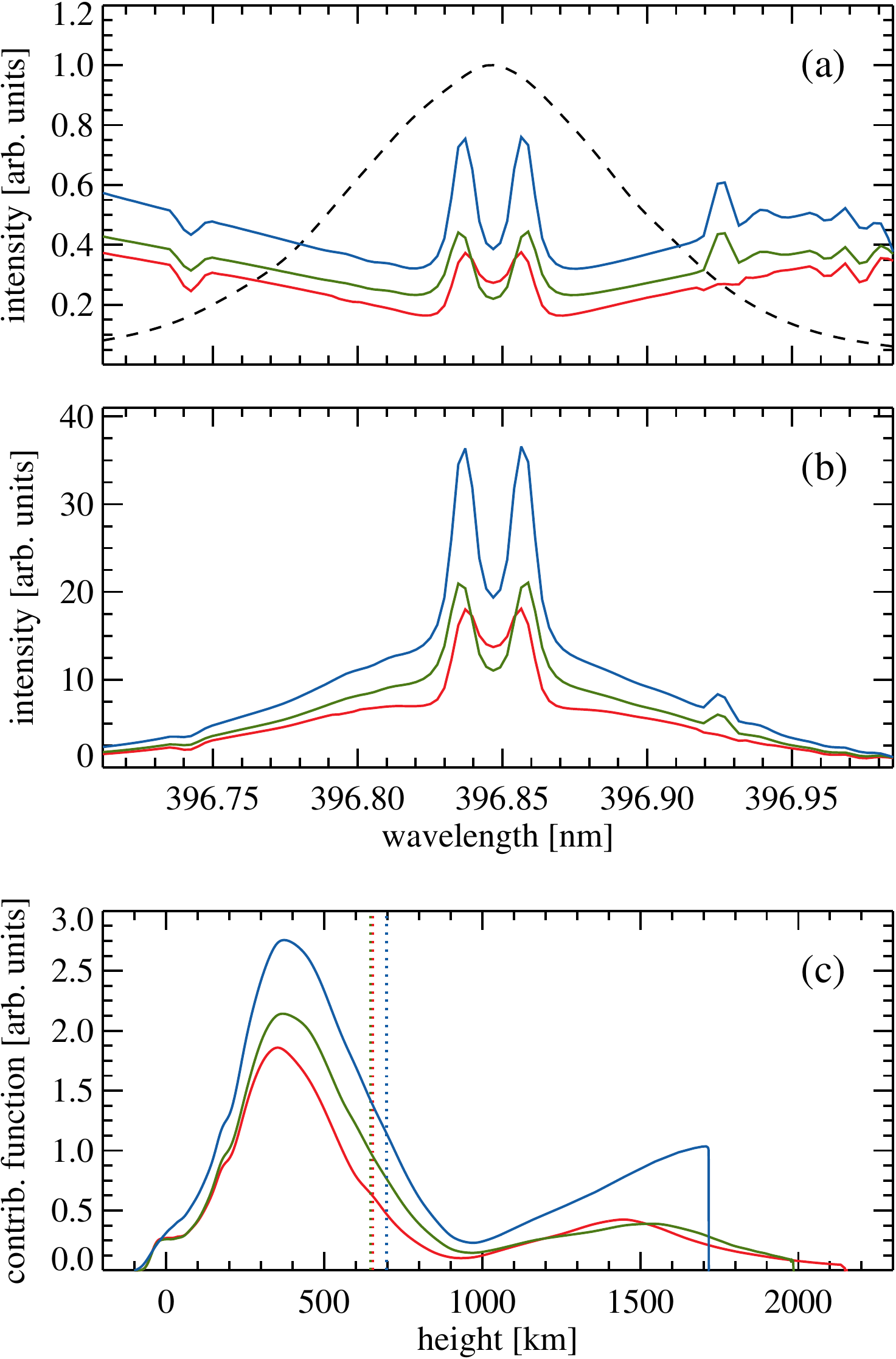}
  \caption{Ca\,{\sc ii}\,H computed from FALC (quiet Sun, red), FALF
  (network, green), and FALP (plage, blue).  Panel (a):
  emergent-intensity spectra.  The dashed curve is the transmission
  profile of the {\sc Sunrise}/SuFI Ca\,{\sc ii}\,H filter.  Panel (b):
  filter-transmitted intensity spectra.  Panel (c): contribution
  functions for the summed transmitted intensities.  The vertical
  lines mark the first-moment height-of-formation averages.}
  \label{fig:FHCaIIH}
\end{figure}

We used three different models from \cite{Fontenla1993}, FALC, FALF,
and FALP, that were constructed to represent average quiet Sun, bright
network, and plage, respectively.
Figure~\ref{fig:FHCaIIH} shows the resulting synthetic Ca\,{\sc ii}\,H
profiles, their multiplication with the transmission profile of the
SuFI Ca\,{\sc ii}\,H filter, and the corresponding CFs for the three
models with average formation heights.
The CFs for FALC and FALF obtain their largest contributions from the
photosphere below 550~km; only FALP predicts a larger fraction
($\approx$60\%) from higher layers.
This difference is already evident in the synthetic profiles in which
the H$_{2V}$ and H$_{2R}$ emission peaks are highest for the plage
model, presumably portraying magnetic heating
\citep[e.g.,][]{Linsky1970,Skumanich1975,Ayres1986,Solanki1991,Fontenla2009}. Note, however, that the heights obtained from Figure~\ref{fig:FHCaIIH} may be overestimated, as the narrowband Ca\,{\sc ii}\,H filter of SuFI may have been somewhat offset relative to the line core, as noted by \citet{Solanki2017}.

\begin{figure}[!tp] \centering
  \includegraphics[width=8cm, trim = 0 0 0 0, clip]{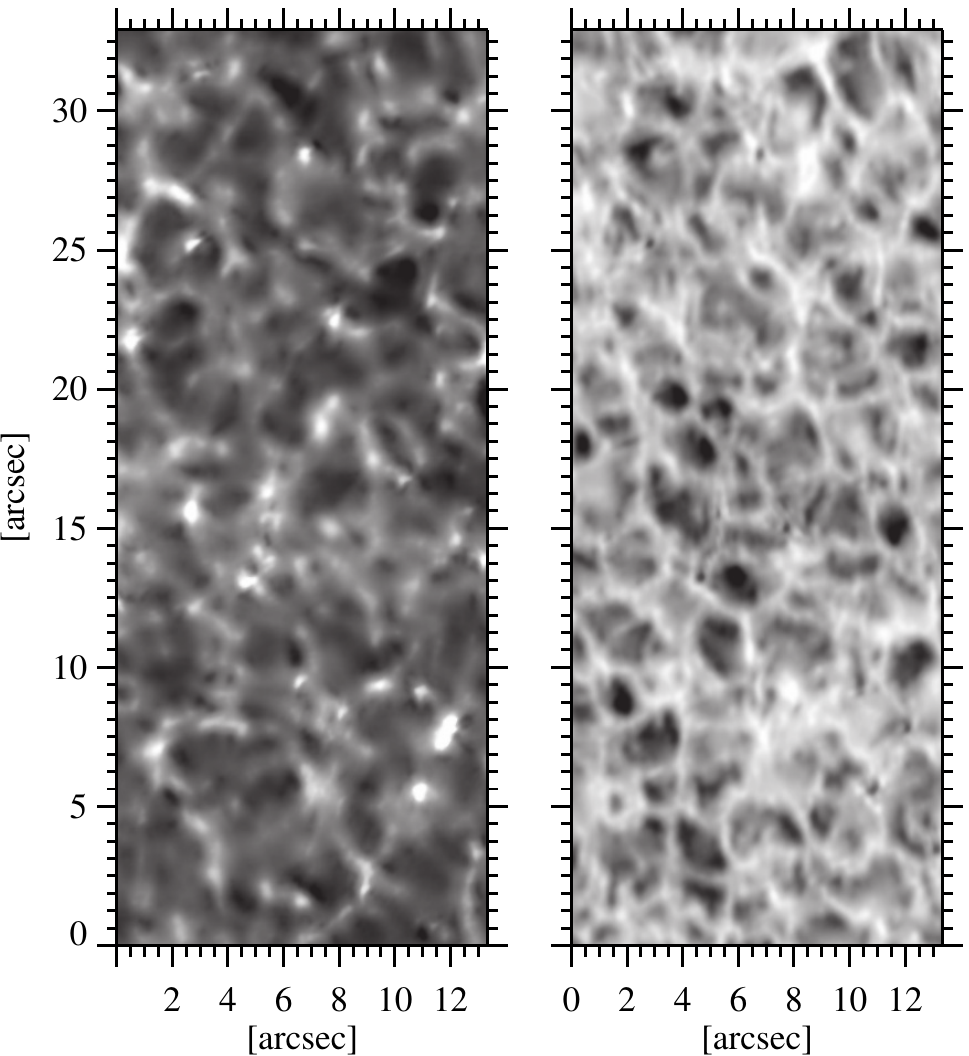}
  \caption[]{Left: quiet area in the narrow SuFI Ca\,{\sc ii}\,H
  passband. 
  Right: corresponding granulation image in the wide 300\,nm passband
  with reversed contrast.}
  \label{fig:obs_qs}
\end{figure}

These standard-model estimates may nonetheless serve to interpret the
background scenes in Figure~\ref{fig:obs_images}. 
We first discuss the quietest areas outlined by the yellow boxes.
For comparison, Figure~\ref{fig:obs_qs} shows a truly quiet area taken
with the Ca\,{\sc ii}\,H filter on 2013 June 13 at 12:53~UT. We comment that these quiet-Sun images were phase diversity reconstructed employing averaged wavefront errors (see \citealt{Hirzberger2010,Hirzberger2011}) and were not processed by MFBD.
Higher resolution of such images from the {\sc Sunrise} flight of 2009 are shown by \citet{Solanki2010}.
The quiet-Sun scene in Figure~\ref{fig:obs_qs} clearly contains no chromospheric features but
only reversed granulation, wave interference patterns including
H$_{2V}$ bright points, and a few magnetic bright points. 
The second panel shows the simultaneous wide-band granulation image
but with intensity reversed so that the now bright intergranular lane
pattern can directly be compared to the cellular ridge patterns in the
first panel.
This shows that the narrowband scene is not simply dominated by
granulation reversal but also contains significant other components. 
A similar study by
\citet{Rutten2004} 
showed that these mostly represent acoustic wave interference,
possibly also gravity-wave interference.
The upshot is that the background patterns represent formation in the
left-hand CF peaks of Figure~\ref{fig:FHCaIIH}, i.e., in the upper
photosphere, and that the SuFI SCFs lie above this height.

\paragraph{Comparison with He~{\sc ii}~30.4~nm}
In order to search for relationships between the SCFs and comparable
elongated fibrilar structures in the upper atmosphere we have compared
our Ca\,{\sc ii}\,H image sequence with corresponding co-temporal and
co-aligned image sequences recorded in He~{\sc ii}~30.4~nm by SDO/AIA.
We take the latter as a proxy for H$\alpha$ fibrils since we have no
simultaneous observations in the latter line and,
at least in active regions, there tends to be general
correspondence between fibrils seen in He~{\sc ii}~30.4~nm and in H$\alpha$
(e.g., the IBIS mosaic
of \citealt{Reardon2012}; 
Figure~1 of \citealt{Rutten2015}). 

However, in our inspections we found no clear one-to-one
correspondence of features in our Ca\,{\sc ii}\,H filtergrams to the
structures in the He~{\sc ii}~30.4~nm images, neither at the same time
nor with a time delay. 
However, even if some of the SCFs (i.e., the longer and thicker ones)
have counterparts in the hotter He~{\sc ii}~30.4~nm line, the spatial
resolution of the AIA images is too low to show very thin fine
structures, such as the SCFs in our SuFI images.

\section{Observation-driven Field Modeling}  \label{sec:modeling}

\subsection{Magnetic Field Extrapolation using a Magnetostatic Approach}
We now discuss the 3D magnetic configuration above the observed solar
region that is obtained from magnetostatic extrapolation of the field
into the chromosphere.
The theory of the used special class of magnetostatic equilibria has been developed in \citet{Low1991} and applied to the quiet Sun in \citet{Wiegelmann2015}.
In this approach, the high spatial resolution {\sc Sunrise}/IMaX vector
magnetogram is embedded into a large-scale and flux-balanced magnetogram of the whole active
region observed by SDO/HMI.
The composite serves as a boundary condition for the magnetostatic
modeling, which yields the field configuration in the non-force-free
layers between the photosphere and the mid-chromosphere more reliably than force-free extrapolations, which are limited to the low plasma-$\beta$ solar corona. 
The original computational box has a FOV of $86\times86$~Mm$^2$ and extends through the solar atmosphere up to a height of 8~Mm (with 200 grid points in the vertical direction).
Deviations from force-free fields occur mainly below 2~Mm in the mixed plasma-$\beta$ upper photosphere and chromosphere.
For more detail, see \citet{Wiegelmann2017}.

\begin{figure*}[!thp] 
\centering
  \includegraphics[width=.96\textwidth, trim = 0 0 0 0, clip]{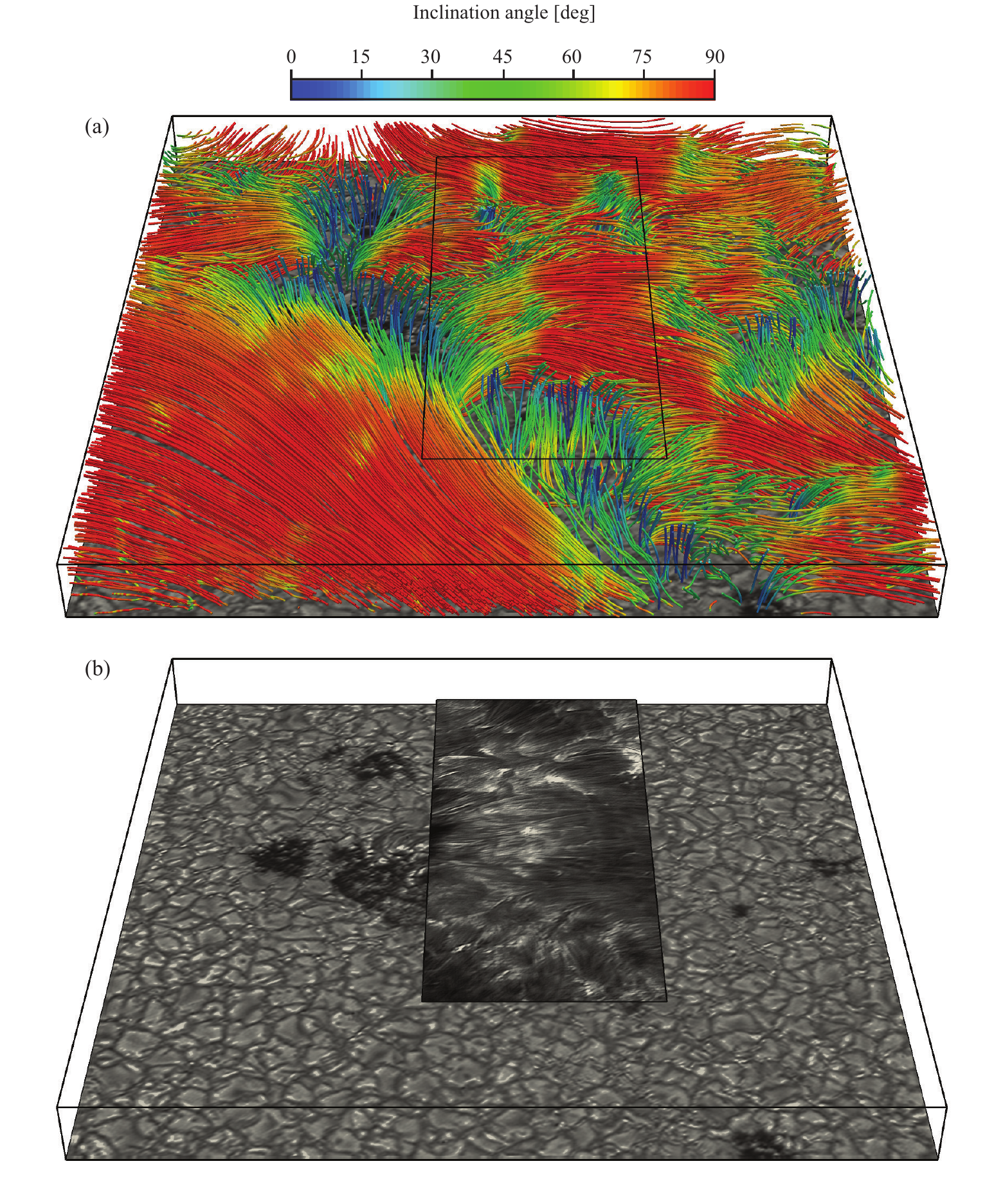}
  \caption[]{Magnetostatic extrapolation of the surface magnetic
  field observed with IMaX and SDO with height.  In panel (a) the
  rectangle specifies the field of view of the SuFI Ca\,{\sc ii}\,H image;
  the colors specify inclination angles.  Panel (b) shows the
  corresponding observed IMaX continuum image and SuFI Ca\,{\sc ii}\,H
  image.  The boxes measure $37\times37$ Mm$^2$ horizontally and
  1400~km vertically, with vertical scale doubling for better
  visibility.} 
  \label{fig:extrapolated1}
\end{figure*}

We limit our displays to the FOV of the {\sc Sunrise}/IMaX images and
to a height of up to only 1400~km for better visualization. 
As in Section~\ref{sec:bifrost} we trace and plot field lines passing
through starting locations that were selected randomly, in this case
within the subvolume above 700\,km, i.e., spanning the 700-1400\,km
height range. 
This selection implies that only field lines reaching heights of
700\,km or higher are shown, i.e., chromospheric fields. One reason for restricting ourselves to these field lines is because (as we saw in Section~\ref{sec:bifrost}) fields peaking lower issue from weaker fields at the solar surface that are more strongly affected by noise in the IMaX data. Hence these higher fields are more reliably extrapolated.  

\begin{figure}[!t]
\begin{minipage}{\textwidth} \centering
  \includegraphics[width=.96\textwidth, trim = 0 0 0 0, clip]{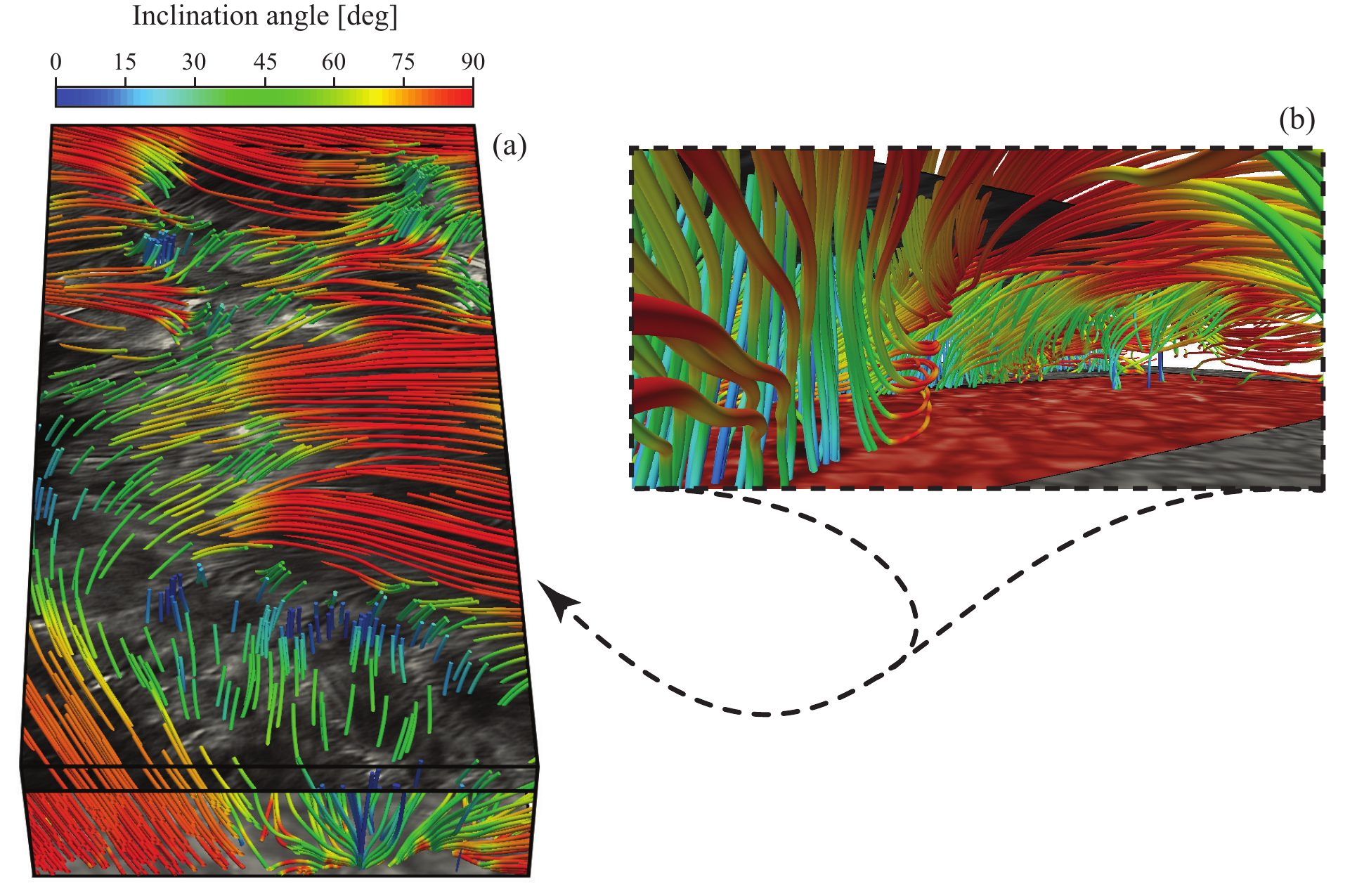}
    \captionof{figure}{Panel (a): same as Figure~\ref{fig:extrapolated1}(a) but limited to the SuFI FOV and with the
  Ca\,{\sc ii}\,H image inserted at 1000~km height. Panel (b)
  gives an inside view of the apparent canopy from a viewing
  angle shown by the arrow pointing to panel (a).  The SuFI FOV is indicated by
  red color on the IMaX continuum intensity image in panel (b).}
  \label{fig:extrapolated2}
\end{minipage}
\end{figure}

Figure~\ref{fig:extrapolated1}a shows this selective field
extrapolation at the moment the images plotted in
Figure~\ref{fig:obs_images} were recorded.
The FOV of the SuFI Ca\,{\sc ii}\,H images is marked by the black
rectangle.
Many field lines form low-lying, rather flat loops that close within the FOV (i.e., those with blue/green at both ends, and long red parts in the middle; examples are field lines shown near the mid-right edge of the FOV of the SuFI images). Some other field lines go straight up (such as those fanning out from the pores, e.g., near the bottom of the SuFI FOV, or outside the SuFI FOV on its mid-to-upper left side), while others remain flat for some distance and then go out of the FOV (unipolar canopy type; examples are the field lines depicted in the upper left and in the lower-right corners of the box). Many other filed lines pass in and out of the displayed FOV at the sides (e.g., those shown close to the lower left corner).
For reference, the IMaX continuum image and the Ca\,{\sc ii}\,H
filtergram are depicted in Figure~\ref{fig:extrapolated1}b. 
The figure covers the full IMaX FOV rather than only the small SuFI FOV
to permit inspection of the surrounding field configuration.
Comparison of the two panels in Figure~\ref{fig:extrapolated1}
indicates that the majority of the SCFs are oriented in the same
directions as the field lines within the SuFI FOV. 
There are, however, disagreements in the lower left corner and near
the upper edge of the Ca\,{\sc ii}\,H image. 
These are the quietest regions in the SuFI images (outlined by
yellow frames in panels (c) and (d) of Figure~\ref{fig:obs_images})
and sample only low atmospheric heights.
Although the extrapolation does contain chromospheric field lines
above these areas there are no SCF-like features with sufficient opacity
for visibility.

Overall, the comparison suggests that the majority of the SCFs do
outline magnetic fields, but with some exceptions.

Field inclinations should carefully be taken into account in this
comparison, since a more vertical field produces shorter SCFs within the
Ca\,{\sc ii}\,H response range in height.
Thus, the presence of relatively long SCFs in the center-right part of
the FOV (right-hand part of the larger ellipse in
Figure~\ref{fig:obs_images}) agrees with the presence of long field
lines lying nearly horizontally in this area, while shorter SCFs elsewhere
indeed correspond to less horizontal field lines.

Figure~\ref{fig:extrapolated2}(a) is similar to figure~\ref{fig:extrapolated1}a, but limited to the FOV of SuFI and with the Ca\,{\sc ii}\,H image inserted at 1000~km height. This facilitates the comparison of the field lines with different inclination angles and the SCFs in the Ca\,{\sc ii}\,H image.
A small 
\vspace*{13.35cm}
\noindent part of the field configuration in
Figure~\ref{fig:extrapolated2}(a) is shown enlarged as a 2D cut in Figure \ref{fig:extrapolated2}(b) from inside the field configuration
along a line of sight, indicated by the arrow in Figure
\ref{fig:extrapolated2}(a).
The SuFI FOV is indicated by red-coloring the IMaX continuum image. 
It illustrates the canopy-like dome made by those field lines that
make it to 700~km or higher and are mostly rooted in the pore and
plage field concentrations.

We comment that these extrapolations are probably missing low-lying fields, because due to the higher noise of the IMaX observations on {\sc Sunrise}-II relative to {\sc Sunrise}-I many weak-field features are missing. From the work of \citet{Wiegelmann2010} and \citet{Wiegelmann2013}, as well as from our study described in Section~\ref{sec:bifrost}, we know that these weak (small-flux) features lie at the footpoints of low-lying loops. Alternatively, the low-lying loops are not well modeled because of, e.g., limitations of the magnetostatic equilibrium.

The conclusion of this section is that the orientation of the
majority of the observed SCFs corresponds well to the field
configuration in the low chromosphere as derived from the IMaX
magnetograms. 
However, a small fraction does not map the extrapolated field lines. 
The latter cases are mostly located in less-organized areas.

\begin{figure*}[!thp] \centering
    \includegraphics[width=.96\textwidth, trim = 0 0 0 0, clip]{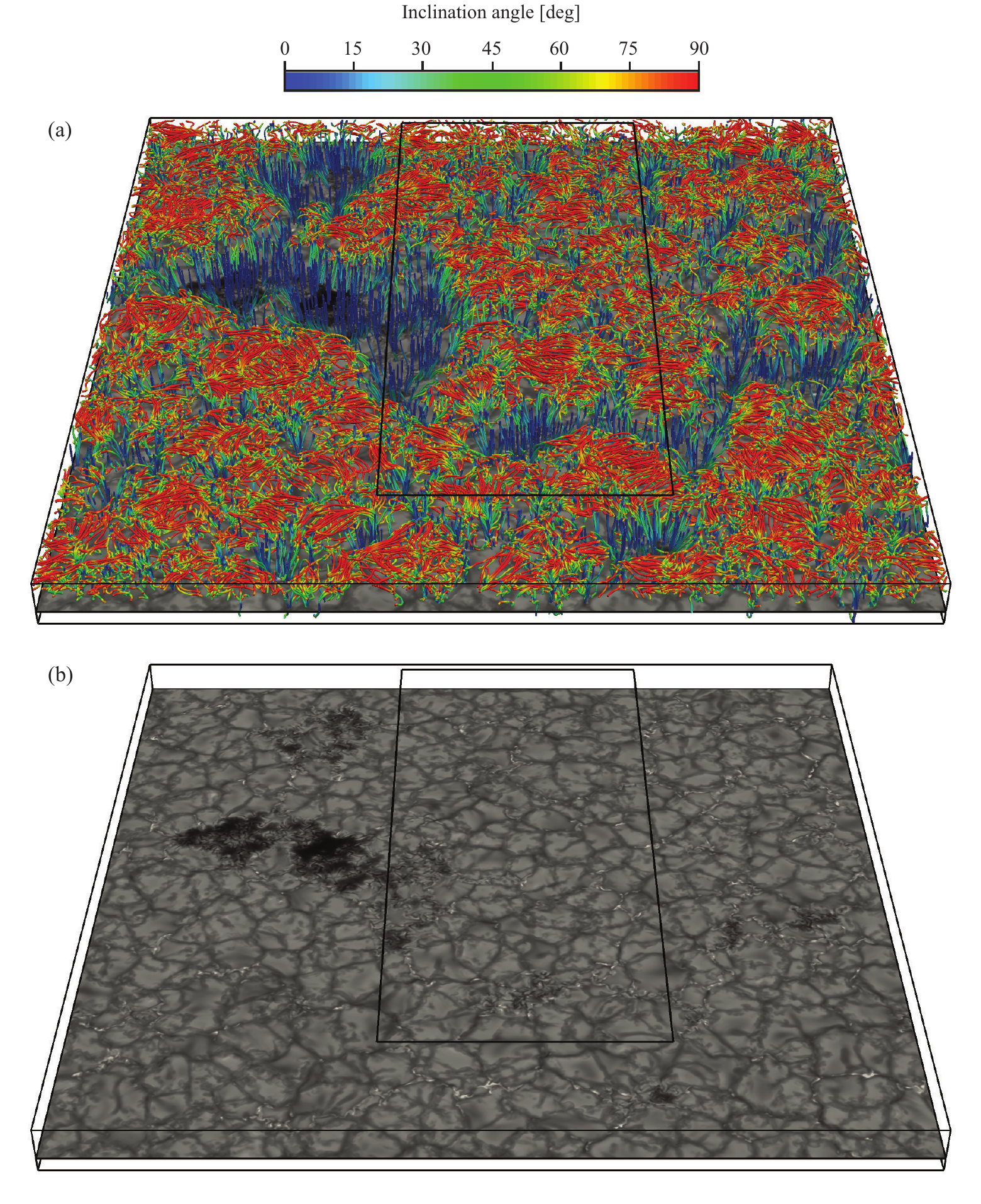}
    \caption{Magnetic field configuration in the solar photosphere from
    a novel MHD-assisted inversion technique (panel a).  The colors
    specify inclination angles.  The synthetic image in panel (b)
    represents a target similar to the IMaX observation in
    Figure~\ref{fig:extrapolated1}(b).  An area with the size of the
    SuFI Ca\,{\sc ii}\,H images is indicated in both panels with a
    rectangle.  The volume measures $34\times34$ Mm$^2$ horizontally
    and 1000~km vertically (of which 700~km lie above the continuum
    intensity level).  The vertical scales are doubled for better
    visibility.}
  \label{fig:masi}
\end{figure*}

\subsection{MHD-assisted Stokes Inversion}
\label{sec:masi}

In addition to our inspection of the magnetic configuration determined
from observed-field extrapolation, we similarly investigate the
photospheric field geometry with a novel MHD-assisted Stokes inversion
technique (MASI; ~\citealt{Riethmuller2017}). 
In this approach, an initial MHD simulation carried out with the MURaM
code \citep{Voegler2005} containing similar features as in the
observed area is used to obtain the 1D atmospheric models providing the best-fit synthetic profiles to the Stokes profiles observed by IMaX in each pixel.
This is done by reshuffling the simulation columns to best
match the observations and then rerunning the simulation to get a
physically consistent result. After multiple iterations this
technique delivers a modified MHD simulation starting from a magnetic
geometry and thermodynamic structure similar to that of the
observation. 
After a simulation run of about 30~minutes solar time to reach
relaxation, a physically consistent MHD simulation is obtained which
closely resembles the observation. 
For more details see \citet{Riethmuller2017}.

We took such a resulting MASI cube and plot in
Figure~\ref{fig:masi} the field configuration in
a similar manner as in Figures~\ref{fig:bifrost},
\ref{fig:extrapolated1} and \ref{fig:extrapolated2}.
It represents heights below 700~km on average from the solar surface.
The area measures $34\times34$~Mm$^2$, similar to (but slightly smaller than) the IMaX FOV; the resolution is 20.8~km/pixel. 
Field lines were again traced starting from random points in a layer reaching from the top boundary to 350~km below it.
A continuum intensity image is added in panel (b) for reference. 
We note that these simulations are based on the assumption of LTE and
are limited to photospheric heights.
Therefore, the field configuration from MASI does not contain
chromospheric structures as the ones defining the majority of our
SCFs. 
Hence, Figure~\ref{fig:masi}(a) represents only photospheric field lines
for an FOV similar to that in our observation.
The black rectangle indicates a simulation area with the same size and
similar content as our SuFI FOV.
It shows many small loops in the upper photosphere. These small loops are likely peaking from weaker fields at the solar surface (as we saw in Section~\ref{sec:bifrost}) compared to the extrapolated fields illustrated in Figures~\ref{fig:extrapolated1} and \ref{fig:extrapolated2}. Such photospheric  loops could not be reliably extrapolated because the weaker fields are more strongly affected by noise in the IMaX data.
Many of the small low-lying loops shown in Figure~\ref{fig:masi}a display a preferred orientation in
the $x$ direction, which agrees well with the observed SCFs and the field lines at a somewhat greater height in the magnetostatic extrapolation. 
Although this simulation result and the observation are not comparable on a one-to-one basis, we speculate that such low short loops may represent some of the shorter observed SCFs.
Remember that the passband of the SuFI Ca\,{\sc ii}\,H filter
yields a wide contribution range from photosphere to chromosphere
(Section~\ref{sec:observations}), so that some of the fibrils may indeed be lying at or slightly below the temperature minimum height. However, they cannot lie too low, as no such fibrillar structures are visible in the line core of the IMaX line, Fe~{\sc i}~525.02~nm.

The upshot from this modeling together with the
magnetostatic extrapolation is, that the observed SCFs map field lines in
the low chromosphere, around 700-1000\,km height.  
\vspace{2mm}
\section{Conclusion}    \label{sec:conclusion}

The SCFs in our Ca\,{\sc ii}\,H images suggest the existence of
small-scale heating events in or near plage and network field
concentrations that produce thin long signatures in the Ca\,{\sc ii}\,H line
that become visible at the bandpass and high angular resolution
of {\sc Sunrise}/SuFI.

Our detailed field extrapolation and MHD-simulation-assisted field
modeling both suggest that the majority of these features are
aligned with and chart magnetic field lines in the low solar
chromosphere at a height of around 700--1000~km. 
Such field lines constitute a low canopy-like dome across the
neighboring quiet areas. MHD simulations suggest that such a
dome is defined by the field strength where the field lines forming it
originate. Areas with less well-organized fields seem to have
less good alignment between field elongated structures in Ca\,{\sc ii}\,H.

Future studies should address the nature of these heating events and
their relation to comparable phenomena as spicules-II and long
H$\alpha$ fibrils, but these then require simultaneous H$\alpha$ and Ca\,{\sc ii}\,H (or Ca\,{\sc ii}\,K)
observations of at least the quality shown here.

\acknowledgements The German contribution to {\sc Sunrise} and its reflight was funded by the Max Planck Foundation, the Strategic Innovations Fund of the President of the Max Planck Society (MPG), DLR, and private donations by supporting members of the Max Planck Society, which is gratefully acknowledged. The Spanish contribution was funded by the Ministerio de Econom\'{\i}a y Competitividad under Projects ESP2013-47349-C6 and ESP2014-56169-C6, partially using European FEDER funds. The HAO contribution was partly funded through NASA grant number NNX13AE95G. This work was partly supported by the BK21 plus program through the National Research Foundation (NRF) funded by the Ministry of Education of Korea. S.J. receives support from the Research Council of Norway. T.W. acknowledges support by DFG-grant WI 3211/4-1. M.S. acknowledges support from the European Research Council under the European Union's Seventh Framework Programme / ERC Grant agreement nr. 291058. The National Solar Observatory (NSO) is operated by the Association of Universities for Research in Astronomy (AURA) Inc. under a cooperative agreement with the National Science Foundation.

\bibliographystyle{aa} \bibliography{Shahin}

\end{document}